\documentclass[aps,twocolumn,prl,10pt]{revtex4-1}

\usepackage{color,colordvi}
\usepackage{amsmath}
\usepackage{amssymb}
\usepackage{graphicx}

\def\be{\begin{equation}}
\def\ee{\end{equation}}
\def\bee{\begin{eqnarray}}
\def\eee{\end{eqnarray}}

\def\kb{k_{\rm B}}

\def\vec#1{{\boldsymbol{#1}}}		
\def\kB{\ensuremath{k_{\rm{B}}}}	

\def\up{\uparrow}
\def\down{\downarrow}

\newcommand{\bbbone}{{\mathchoice {\rm 1\mskip -4mu l}{\rm 1\mskip 
-4mu l}{\rm 1\mskip -4.5mu l}{\rm 1\mskip -5mu l}}}

\begin{document}
 
\title{Cooling a magnetic nanoisland by spin-polarized currents}
\author{J. Br\"uggemann$^{1}$, S. Weiss$^{2}$, P. Nalbach$^{1}$, and M.
Thorwart$^{1}$ }
\affiliation{$^{1}$ I. Institut f\"ur Theoretische Physik, Universit\"at
Hamburg, Jungiusstra\ss e 9, 20355 Hamburg, Germany\\
$^{2}$ Theoretische Physik, Universit\"at Duisburg-Essen \& CENIDE, 47048
Duisburg, Germany}

\date{\today}

\begin{abstract}
We investigate cooling of a vibrational mode of a magnetic quantum dot by a
spin-polarized tunneling charge current, exploiting the
magnetomechanical coupling. The spin-polarized current polarizes the
magnetic nanoisland, thereby lowering its magnetic energy. At the same time,
Ohmic heating increases the vibrational energy. 
A small magnetomechanical coupling then permits to remove energy from the
vibrational motion and cooling is possible. We find a reduction of the
vibrational energy below $50\%$ of its equilibrium value. The lowest 
vibration temperature is achieved for a weak electron-vibration coupling and a
comparable magnetomechanical coupling. The cooling rate 
increases at first with the magnetomechanical coupling and then saturates.  
\end{abstract}

\maketitle

\paragraph*{} 

The ongoing miniaturization of electronic devices will sooner or later have to
face the problem of how to efficiently remove inevitable heating from the
devices. Up to now, essentially all electronic devices operate on the basis of
dumping heat via passive sinks via the supporting structure. Active or
dynamical nano-cooling has received little attention, although passive thermal
transport is inefficient at the nanoscale. In addition, most present
nano-electronic or -magnetic devices function at low temperatures only.
Efficient applications will thus demand dynamic nanorefrigerators. Nano-cooling
would also facilitate new experiments which are not conceivable today due to the
spacious equipment required for cooling. 

Various forms of nanorefrigerators in which heat is carried by an electronic
\textit{charge} current have been proposed \cite{Pekola}. We, in contrast,
propose to use the electronic \textit{spin} for 
cooling, similar to the macroscale magnetocaloric demagnetization cooling
\cite{Pobell2007}. Typically, cooling requires to open and close heat links
which is realized at the macroscale by mechanically moving parts or using
coolants. This is impractical at the nanoscale. Instead, we propose to use
spin-polarized currents to polarize a magnetic nanoisland, thereby lowering its
magnetic energy. Subsequent energy exchange due to a
magnetomechanical coupling between the magnetic and
vibrational degrees of freedom can then reduce the vibrational energy. Electric
losses give rise to Ohmic heating. Thus, a net
cooling is reached when the polarization of the magnetic moment is
faster than Ohmic heating. 

\begin{figure}
\includegraphics[width=.92\linewidth]{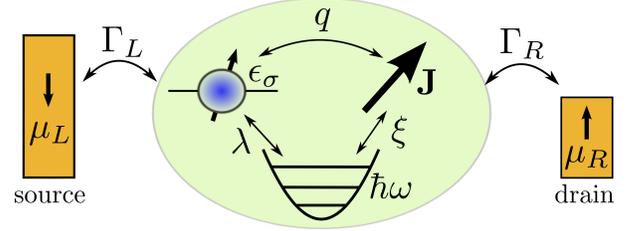}
\caption{\label{fig1}
Schematic consisting of ferromagnetic leads, a quantum dot
with a localized magnetic moment and a single vibrational
mode.}
\end{figure}

In detail, we investigate the nonequilibrium quantum dynamics of a magnetic
quantum dot with a single electronic level, a local magnetic moment $\vec{J}$
and a single vibrational mode as sketched in Fig. \ref{fig1}. The magnetic 
quantum dot is
weakly coupled to ferromagnetic spin-polarized electronic leads via tunneling.
Charge and spin currents through quantum dots with an additional magnetic moment
{\em or\/} an additional vibrational mode have been studied extensively
\cite{Flens03a,Flens03b,Nitzan06,Efros01,Koenig10,Wegewijs11,Koenig,Launay2013},
but the
full model including both has been unexplored so far. For weak tunneling
contacts, the sequential tunneling processes dominate and a description by 
classical or quantum master equations is adequate
\cite{Schoeller96,Koenig0}. Spin-polarized currents can polarize the local
magnetic moment $\vec{J}$ against an applied magnetic field and
thus can lower its energy. At the same time, the vibrational motion is heated 
due to Ohmic losses. 
The final goal is to cool the vibrational mode by means of an effective
magnetomechanical coupling which  allows for an energy
exchange between the local magnetic moment $\vec{J}$ and the vibrational mode.
Such magnetomechanical couplings
 have been suggested theoretically for a nanomechanical cantilever which
interacts with a ferromagnetic tip on its surface \cite{Bargatin03}, for a
nitrogen-vacancy (NV) impurity in diamond which
couples to the magnetic tip of a nanomechanical cantilever \cite{Rabl2009}, for
a magnetic nanoparticle or a single molecular magnet attached to a torsional
doubly clamped resonator \cite{Kovalev2011}, and for a single electron spin
which couples to a flexural mode of a suspended carbon nanotube
\cite{Palyi2012}. The experimental detection of the coupling of an individual
electron spin and the magnetic tip of a cantilever was successful
\cite{Rugar04}. Magnetomechanical coupling of a single NV
center to a SiC cantilever has also been demonstrated 
\cite{Arcizet2011,Kolkowitz2012}.
Furthermore, a strong coupling between the magnetic moment of a
single molecular magnet and a suspended carbon nanotube has been
reported \cite{Ganzhorn2013}. This magnetomechanical coupling can be
explored for {\em cooling\/} the vibrational motion. We
provide an intuitive picture of the concept and determine viable parameters for
the cooling of the vibration \textit{below} its equilibrium 
temperature. The lowest vibrational energy is achieved for a weak
electron-vibration coupling and a comparable magnetomechanical coupling. 

\paragraph*{Model}
Our minimal model is sketched in Fig.\ref{fig1}. The quantum dot
is given as a single electronic level with energy $\epsilon_{0}$. For small
dots, the local charging energy exceeds all other energies, and
a double dot occupancy is forbidden. The Hamiltonian is
$H_{d}=\epsilon_{0}\left(a^{\dag}_{\uparrow}a_{\uparrow}+a^{\dag}_{\downarrow}a_
{\downarrow}\right)+\frac{g\mu_B}{\hbar} B s_z$, with electron operators
$a_{\sigma}$ and $a^{\dag}_{\sigma}$, the g-factor
$g$, and the Bohr magneton $\mu_B$. The spin projection quantum number is
$\sigma$ and $s_z$ the $z$-component of the electron spin $\vec{s} =
\frac{\hbar}{2}\sum_{\sigma,\sigma'}a^\dagger_\sigma {\boldsymbol
\sigma}_{\sigma\sigma'} a_{\sigma'}$. A small external field $B$ splits the spin
states along the quantization axis. The localized magnetic moment is modelled by
a spin-$1/2$ impurity. Generalization to higher spin values is
possible. We denote with $J_{z}=\pm \hbar/2$ the projection onto the
quantization axis. The corresponding Hamiltonian is
$H_{J}=\frac{g\mu_B}{\hbar}BJ_z + \frac{q}{\hbar^2}(\vec{s}\cdot\vec{J})$,
where
the electronic spin and the local magnetic moment $\vec{J}$ are coupled by
an exchange interaction of strength $q$, for simplicity, assumed to be
isotropic. To study the dynamical
heating, a vibrational mode of frequency $\omega$ is coupled to the  
dot. Using bosonic ladder operators $b$ and $b^{\dag}$,
we obtain $H_{ph}=\hbar\omega b^\dagger b +
\lambda\left(b+b^\dagger\right)\sum_\sigma a_\sigma^\dagger a_\sigma$. 
The linear coupling of the electronic occupation to $(b+b^\dagger)$ allows to
excite or relax the vibrational mode by each tunneling electron. It also 
describes  heating of the mode due to charge current.  
The crucial ingredient is a magnetomechanical coupling between
the magnetic moment $\vec{J}$ and the vibrational degree of freedom included by
\begin{equation}\label{eqxi}
H_{J-\rm{ph}} = \frac{\xi}{\hbar} \left(b+b^\dagger\right) \left(J_+ +
J_-\right)\, , 
\end{equation}
where $J_\pm = (J_x \pm i J_y)/2$ are spin-$1/2$ ladder operators inducing
transitions between the magnetic states  when at the
same time the vibrational mode changes its angular
momentum state \cite{Spinphonon}.
Experimental values for $\xi$ and $\lambda$ can readily be
extracted for the existing experimental set-ups. Ganzhorn et al.\
\cite{Ganzhorn2013} have realized a set-up with a single molecular
magnet covalently bound to a carbon nanotube suspended between two
leads. The molecule has a magnetic ground state of $|\vec{J}|=6$ and the
groundstate doublet $J_z=\pm 6$ is separated from the excited states by several
hundreds of Kelvin. Hence, an
Ising-like spin flip between these two states via quantum spin
tunneling is dominant and the physics is well described by an effective 
spin-$1/2$. The spin flip is accompanied by a transition in the
vibrational mode and one finds \cite{Ganzhorn2013} $\omega=34$
GHz,
$\xi=1.5$ MHz, and $\lambda=\omega\sqrt{g}=26$ GHz for $g=0.6$, implying that
$\xi/\omega \sim 4 \times 10^{-5}$ and $\lambda/\omega \sim 0.76$. For the NV
centers \cite{Arcizet2011}, we have $\omega=2 \pi \times 625$ kHz and 
$\xi=g\mu_B \frac{\partial B}{\partial z}\sqrt{\hbar/(2m\omega)}\sim 172$ Hz,
implying that $\xi/\omega \sim 4 \times 10^{-6}$. 
Another realization \cite{Kolkowitz2012} yields $\omega=2
\pi \times 80$ kHz and $\xi\sim 8$ Hz, such that $\xi/\omega \sim 
10^{-4}$.
The quantum dot is tunnel coupled to two ferromagnetic (FM) leads, whose
magnetization  directions  are in general noncollinear. They are
modelled as non-interacting  electron reservoirs $H_{\rm{leads}} =
\sum_{k\alpha} \left(\epsilon_{k\alpha}-\mu_\alpha\right)
\left(c_{k\alpha+}^\dagger
c_{k\alpha+} + c_{k\alpha-}^\dagger c_{k\alpha-}\right)$, where $c_{k\alpha
\pm}$  represents the annihilation operator for an
electron with the wave number $k$ and the majority/minority spin in the lead
$\alpha=L,R$ and $\mu_{L/R} = \pm eV/2$ is the chemical potential of the leads
shifted by the applied bias voltage $V$. In  the FM 
leads, the spin species have different density of states at the Fermi
energy. We define the polarization $p_\alpha =
(\nu_{\alpha,+}-\nu_{\alpha,-})/(\nu_{\alpha,+}+\nu_{\alpha,-})$ of lead
$\alpha$ by
the relative difference in the density of states $\nu_{\alpha,\pm}$ for majority
/ minority spins at the Fermi energy. We use
$p_L=p_R=p$. All energies
are measured relative to the Fermi energy at zero polarization. Spin
dynamical effects, which affect the vibrational dynamics, are
influenced by the relative angle between the magnetization
directions of the leads due to the exchange field on the dot
\cite{Koenig}. The source lead polarization is chosen as antiparallel to 
 $B$. We consider three set-ups with drain polarization parallel
($\downarrow \downarrow$), perpendicular ($\downarrow \rightarrow$) or
anti-parallel ($\downarrow\uparrow$) to the source. To have an overall
quantization axis, the tunneling Hamiltonian depends explicitly on spin rotation
matrices as $H_{t} = \sum_{k,\alpha=L/R} \left[t_{k\alpha} A_\mu
\Lambda_{\mu\nu}^{(\alpha)}
C^\dagger_{\nu ;k,\alpha} + \rm{h.c.}\right] $
with $A=(a_\up,a_\down)$, $C_{k,\alpha}=(c_{k\alpha +},c_{k\alpha -})$ and
$\Lambda^{(L)}=\Lambda^{(R,para)}=\bbbone$, $\Lambda^{(R,anti)}=\sigma_x$ and
$\Lambda^{(R,perp)}=(\bbbone-i\sigma_y)/\sqrt{2}$ for the three setups. The
hybridization with the dot state in the wide-band limit is given by
$\Gamma_\alpha = 2\pi |t_{k\alpha}|^2(\nu_{\alpha,+}+\nu_{\alpha,-})$. 

\paragraph*{Method}
 Spin polarization and excitations of the vibration are explicit dynamical
processes while 
the lead electrons can be integrated out. The time
evolution of the reduced density matrix includes the 
dot electrons, the magnetization and the vibration and obeys the kinetic
equation
\cite{Schoeller96,Koenig0}
\begin{eqnarray}
\partial_t \rho^{\chi_1}_{\chi_2}(t) &=& -i\left(\epsilon_{\chi_1} -
\epsilon_{\chi_2}\right) \rho^{\chi_1}_{\chi_2}(t) \nonumber \\ 
&& -\int_{t_0}^t dt' \sum_{\chi_1'\chi_2'} M^{\chi_1'
\chi_1}_{\chi_2' \chi_2}(t,t') \rho^{\chi_1'}_{\chi_2'}(t').
\end{eqnarray}
with $\epsilon_{\chi_{i}}$ describing  the system's eigenenergies. It includes
all quantum coherences of the system as well as all 
nonequilibrium effects due to the leads. In practice, we diagonalize numerically
the Hamiltonian of dimension $6n$ spanned by the states spin-up, spin-down and
empty for the dot electron level with spin-up or spin-down of the local
magnetization. $n$ is the number of vibrational states  necessary for
convergence and depends on $\lambda, T$ and $V$. Here, $n=6$ is
sufficient.
For small tunnel couplings 
$\Gamma_{\alpha}\ll (B, q,\lambda,\kb T)$, the tensor $M$ can be expanded to
lowest 
orders in $\Gamma_{\alpha}$. Taking into account the finite bias voltage, this
gives irreducible self energies on the Keldysh contour
\cite{Schoeller96,Koenig0}. The leads are held at temperature $T$.

In the Markov limit, we solve the kinetic equation and calculate the occupation
probabilities for the subspaces of the dot, the local magnetic moment and
vibration are by further tracing over the respective other degrees
of freedom. Then, we monitor the spin polarization of the dot and 
population of  vibrational states. The 
electron current follows by standard means \cite{Koenig}.

\begin{figure}
\includegraphics[width=.48\textwidth]{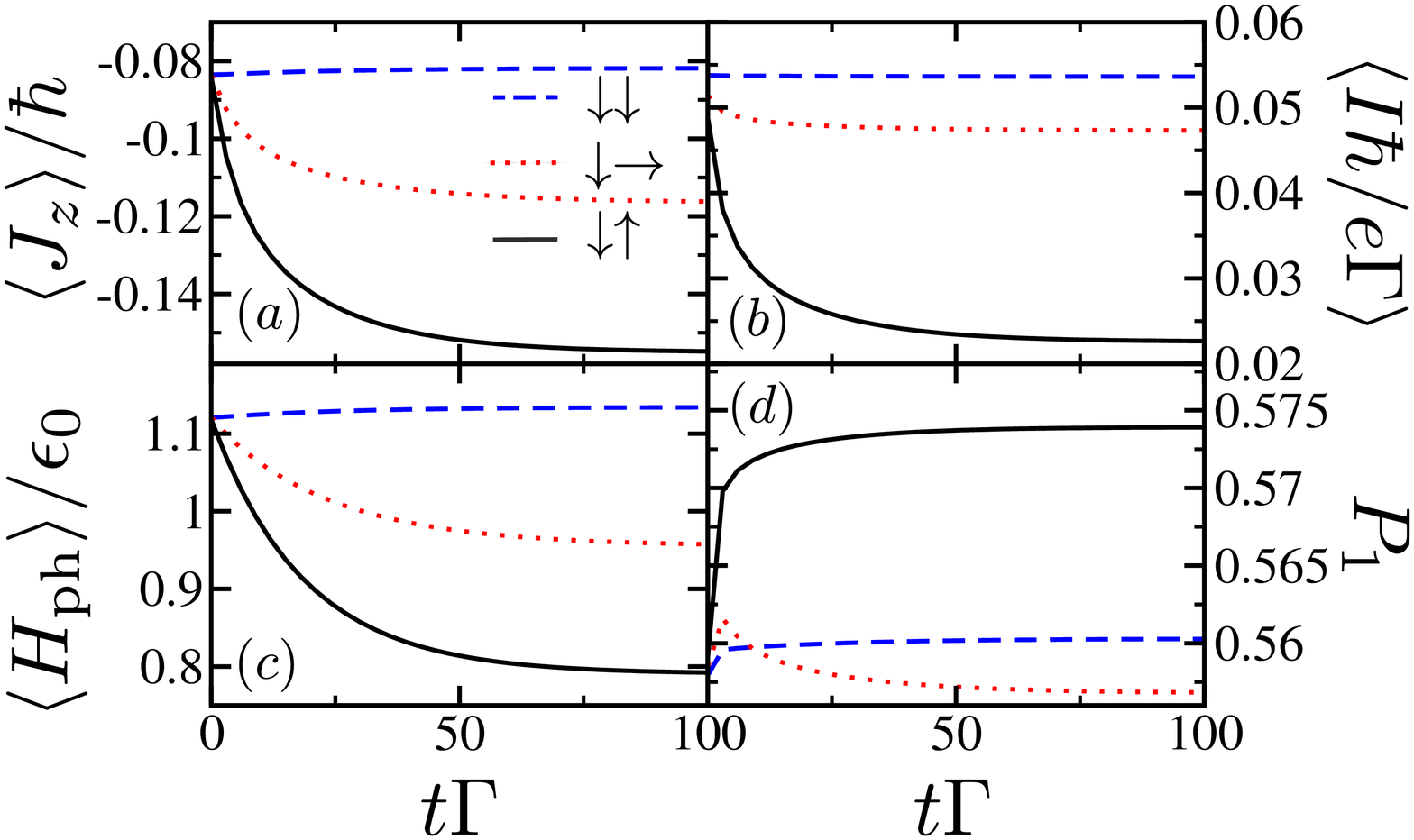}
\caption{\label{fig2}
(a) Polarization $\langle J_z \rangle/\hbar$ of the localized magnetic moment
along the $z$-direction,
(b) charge current $\langle I\hbar/e\Gamma\rangle$, 
(c) Vibrational energy $\langle H_{\rm{ph}}\rangle/\epsilon_0$ and
(d) probability $P_1$ to find one electron on the quantum dot as a function of
time
for the three different lead setups.
The remaining parameters are $p = 0.9$, $\hbar\Gamma=0.01\, \epsilon_0$, $B =
0.7\,\epsilon_0/(g\mu_B)$, $\hbar\omega = 0.75\,\epsilon_0$, $\kB T =
2\,\epsilon_0$, $q = 0.4\,\epsilon_0$, $\xi=0.06\,\epsilon_0$, $\lambda =
0.2\,\epsilon_0$, and $eV = 1.2\,\epsilon_0$.}
\end{figure}

\paragraph*{Principle mechanism}

To provide a qualitative understanding, we  
start from an empty dot and the local magnetic moment $\vec{J}$
being aligned with the magnetic field, say, pointing upwards, i.e., being in the
high-energy state. The vibrational degree of freedom is
assumed to be in a thermal state at the lead temperature $T$. A source lead 
perfectly polarized down enforces all electrons tunneling 
into the quantum dot to have spin-down. The tunneling of a single spin-down
electron onto the dot lowers the dot energy
 because of the $\vec{s}\cdot \vec{J}$-exchange coupling. Due to the FM
leads together with the local magnetic field the electron spin and the localized
magnetic moment are not conserved. This permits a spin flip  of 
the electron spin upwards and the local moment downwards. This lowers the
energy of the local magnetization which is transferred to the electron. If the
drain is
polarized antiparallel to the source, i.e., upwards, the electron can
tunnel preferably out of the dot only after its spin has flipped. 
Thus, antiparallel polarized leads ensure spin-flips and thus lower the total
current.  
Any electronic population of the dot generates vibrational transitions 
 due to the electron-vibration coupling $\lambda$. At finite bias voltage, this
generates Ohmic heating of the vibration above the lead temperature. 
In turn, the magnetomechanical coupling $\xi$ now provides a possibility
for an energy exchange between the magnetization and the vibration. In
particular, the vibration can relax while the
magnetization is flipped upwards and thus parallel to the magnetic field again, 
which reflects its high-energy state. 
The next spin-down electron, which tunnels into the quantum dot,
will remove this energy and a net cooling of the vibrational motion results. 
Non-perfect lead polarizations will lower the cooling efficiency 
since not every electron spin will be forced to flip.

\begin{figure}
  \includegraphics[width=0.49\textwidth]{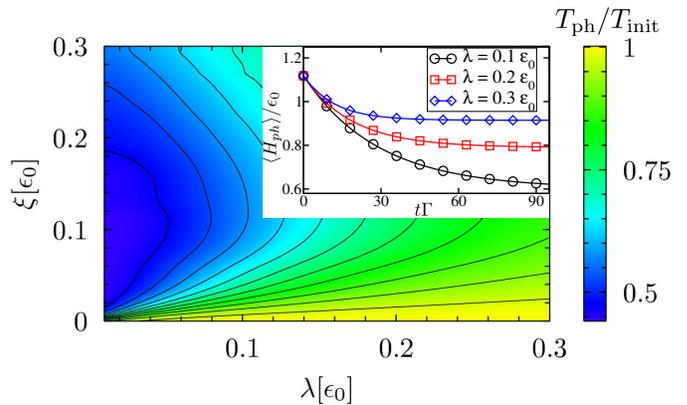}\hfill
\caption{Effective vibrational temperature $T_{\rm{ph}}$ in the stationary limit
versus electron-vibration coupling $\lambda$ and magnetization-vibration
coupling $\xi$.
Inset: Vibrational energy $\langle H_{ph}\rangle$ vs. time for
different electron-vibration coupling strengths $\lambda$. The remaining
parameters 
are as in Fig.\ \ref{fig2}.}
\label{fig3}
\end{figure}

\paragraph*{Results}
To confirm this qualitative picture, we show in Fig.\ \ref{fig2} (a) the time
evolution of the local magnetization $\langle
J_z \rangle$ for the parallel ($\downarrow \downarrow$), the
perpendicular ($\downarrow \rightarrow$) and the anti-parallel
($\downarrow\uparrow$) lead polarizations. The system is
initialized by allowing equilibration with the leads at $V=0$.
A Boltzmann distribution  defines an initial temperature
$T_{\rm init} = \langle H_{\rm{ph}}(t=0)\rangle/\kB$ of the equilibrated
vibration
before applying a bias voltage. For ($\downarrow \downarrow$) alignment,
electrons with
majority spin can tunnel through the
quantum dot without flipping the spin. No spin blockade occurs and the steady
state with a large charge current is reached on a time scale of $\Gamma^{-1}$
(see Fig.\ \ref{fig2} (b)). The energy of the vibrational mode, i.e., $\langle
H_{\rm ph} \rangle/\epsilon_0$, increases slightly, see Fig.\ \ref{fig2} (c).
Subsequently, due to the weak coupling between the vibration and $\vec{J}$, a
slow decrease of the polarization $\langle J_z
\rangle/\hbar$ follows, see Fig.\ \ref{fig2} (a). 
\begin{figure}
 \includegraphics[width=.49\textwidth]{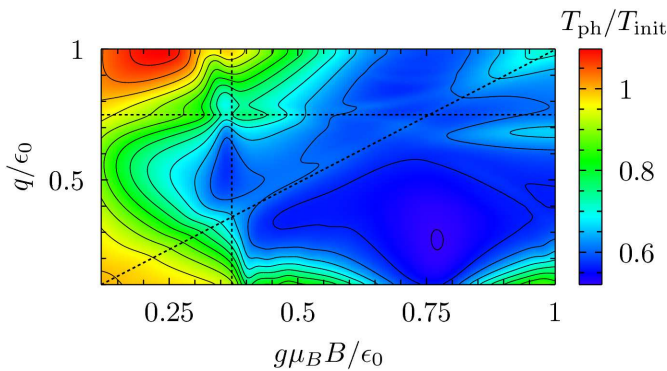}
\caption{Effective vibrational temperature $T_{\rm{ph}}$ in the stationary limit
versus magnetic field $g\mu_B B$ and spin-magnetization coupling $q$. 
The used parameters are $\xi = 0.12\,\epsilon_0$,
$\lambda = 0.1\,\epsilon_0$, with the remaining parameters as in Fig.\
\ref{fig2}.}
\label{fig4}
\end{figure}

For the ($\downarrow \rightarrow$) alignment, the
different spin carrier distributions in
the leads cause a strong  spin blockade  which suppresses the current
(see Fig.\ \ref{fig2} (b)) as compared to the ($\downarrow
\downarrow$) case. At
the same time, a sizeable electron spin accumulates on the dot. A partial
polarization $\langle J_z \rangle/\hbar$ and
cooling, i.e., a decrease of the vibrational energy, is
observed. An
optimal polarization of the local magnetic moment is, however, obtained for the
($\downarrow \uparrow$) alignment. Here, the
minimal steady state current and the lowest energy in the vibrational mode, the
strongest cooling, is found. In
Fig.\ \ref{fig2} (d), the dot occupation probability $P_1$ is shown as
a function of time for all three setups.
The finite charge accumulation for all lead polarizations combined with the
different steady-state currents (see Fig \ref{fig2} (b))
demonstrates the current blockade for the anti-parallel setup. As expected, the
$\downarrow \uparrow$-setup is shown to be optimal in two respects. First, it
produces the largest magnetic polarization of the dot against an applied
magnetic field and, thus, lowers the magnetic energy of $\vec{J}$ optimally.
Second, the spin blockade suppresses the charge current and enforces that each
transmitted electron can contribute to the polarization process and, thus, to
cooling. Additionally, the Ohmic heating is directly connected to the charge
current and therefore suppressed which significantly improves the cooling
efficiency. 

We focus on weak-to-intermediate electron-vibration coupling where the 
vibrational
blockade is absent. For $\xi=0$, we observe for all set-ups an increasing energy
$\langle H_{\rm
ph} \rangle/\epsilon_0$ of the vibrational mode with a stronger heating for a
larger electron-vibration
coupling. For a finite magnetomechanical coupling,  $\xi\neq0$, energy is
exchanged with
the polarized/cooled local moment and a net cooling of the vibrational
mode results. 
In total, these results illustrate the proof of principle of
cooling a magnetic nanodevice by a spin-polarized current.

In order to quantify the cooling, we define an effective vibrational temperature
$T_{\rm{ph}}$ assuming a Boltzmann distribution in the steady state
\cite{phonontemp} as 
\be\label{eq1}
T_{\rm{ph}} = \langle H_{\rm ph} \rangle_{\rm stat}/\kb .
\ee
Fig.\ \ref{fig3} shows the ratio of $T_{\rm{ph}}$ and the
initial temperature $T_{\rm init}$ as a function of the electron-vibration
coupling $\lambda$ and the magnetomechanical coupling $\xi$. Cooling is achieved
in the full parameter regime depicted. As expected, with increasing 
electron-vibration coupling $\lambda$, the effective vibrational temperature 
increases (see inset Fig.\ \ref{fig3}). 
Surprisingly, for fixed $\lambda$, we observe a nonmonotonic dependence of the
cooling as a function of $\xi$ (vertical cut in Fig.\ \ref{fig3}). At first,
$T_{ph}$ decreases with increasing $\xi$. However, a minimum is reached where
 cooling is optimal. For further increasing $\xi$, the
effective vibrational temperature increases again. In Fig.\ \ref{fig4}, $T_{ph}$
is shown as a function of the magnetic field $g\mu_B B$ and the
spin-magnetization exchange coupling $q$. For higher magnetic fields, the
cooling effect is increased since the energy gain due to the spin polarization
is proportional to the magnetic field. 
Additional fine structures are observed and traced back to resonances between
spin flips and vibrational transitions (see dashed lines in Fig.\
\ref{fig4} for the noninteracting limit, 
$\xi,\lambda \to 0$).
\begin{figure}
  \includegraphics[width=0.49\textwidth]{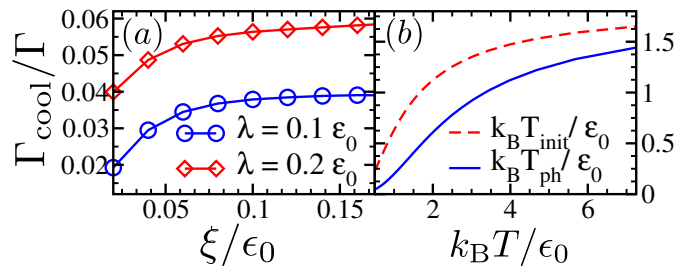}
\caption{(a) Vibrational cooling rate $\Gamma_{\rm cool}$ versus the 
magnetomechanical coupling $\xi$ for two electron-vibration couplings $\lambda =
0.1\,\epsilon_0$ and $\lambda = 0.2\,\epsilon_0$. (b) Effective vibrational 
temperature in the initial, thermalized state ($T_{\rm{init}}$, $V=0$, red
dashed line) and in the final steady state ($T_{\rm{ph}}$, $V>0$, blue solid
line) as a function of the lead temperature $T$ for $\xi = 0.12\,\epsilon_0$ and
$\lambda = 0.1\,\epsilon_0$. The remaining parameters are as in Fig.\
\ref{fig2}.}
\label{fig5}
\end{figure}
An effective temperature as in Eq.\ (\ref{eq1}) can be defined at all times. We
find an exponential approach to the steady state. Thus, we can determine
the effective cooling rate $\Gamma_{\rm
cool}$ by a fit to our numerical results according to $T_{\rm{ph}}(t) \simeq
T_{\rm{ph}}(\infty) +
e^{-\Gamma_{\rm{cool}}t}(T_{\rm{init}}-T_{\rm{ph}}(\infty))$. Fig.\ \ref{fig5}
(a) depicts the cooling rate versus the magnetomechanical coupling for two
values
of electron-vibration coupling, i.e., $\lambda = 0.1\,\epsilon_0$ and $\lambda =
0.2\,\epsilon_0$. We observe initially a strong increase of the cooling rate
with increasing $\xi$. For $\xi\gtrsim 0.1$, however,
the cooling rate saturates.

The initial and the asymptotic effective vibrational temperature as a function
of the lead
temperature $T$ are shown in Fig.\ \ref{fig5} (b). For the depicted
temperature range, the initial temperature $T_{\rm{init}}$ exceeds the
steady state value $T_{\rm{ph}}$, thus maintaining the cooling effect for a
range of lead temperatures of at least one order of magnitude. We also observe
that $T_{\rm{init}}$ is not directly
proportional to the lead temperature which originates from the
preparation of our initial state which includes all couplings. 

\paragraph*{Conclusion}
We have established a simple model to illustrate the principle of cooling a
magnetic nanoisland by a spin-polarized charge current. It is based on the
polarization of the island magnetization by the flowing polarized electron spins
and a subsequent removal of thermal energy from the vibration via a
magnetomechanical coupling. Interestingly, this coupling also overcompensates
Ohmic heating and leads to a significant lowering of the energy stored in the
vibration. We also find that the cooling rate saturates as a function of the
magnetomechanical coupling, implying that not very strong couplings are required
to observe the proposed effect. We are confident that this mechanism could be
realized in magnetic molecular nanojunctions by present day technology. 

We acknowledge financial support by the DFG via the Schwerpunktprogramm Spin
Caloric Transport (SPP 1538). P.N. thanks the BSH Bosch und Siemens
Hausger{\"a}te GmbH for interesting and useful insights into cooling technology.


\begin{thebibliography}{99}
\bibitem{Spinphonon}Electron-spin vibration interactions are absorbed in the
effective spin-independent electron-vibration coupling $\lambda$ and do not
induce spin flips along with vibrational relaxation in our model.
\bibitem{phonontemp} The results do not change significantly when the
various interacting terms are included in this definition.
\end{thebibliography}
\end{document}